\documentclass[twocolumn]{aastex631}
\usepackage{amsmath}

\newcommand{\chan}{\textit{Chandra}}

\newcommand{\xmm}{\textit{XMM-Newton}}

\newcommand{\nustar}{\textit{NuSTAR}}

\newcommand{\src}{G54.1$+$0.3}
\newcommand{\srcp}{PSR J1930$+$1852}
 
\newcommand{\lhaasosrc}{1LHAASO J1929+1846u}

\begin{document}
\title{\nustar\ and \xmm \  Observations of \srcp \ and Its Pulsar Wind Nebula}

\correspondingauthor{Jason Alford}
\email{alford@nyu.edu}

\author[0000-0002-2312-8539]{J. A. J. Alford}
\affil{Division of Science, NYU Abu Dhabi, PO Box 129188, Abu Dhabi, UAE}
\affil{Center for Astrophysics and Space Science (CASS), NYU Abu Dhabi, PO Box 129188, Abu Dhabi, UAE}
\author[0000-0001-8630-5435]{G.-B. Zhang}
\affil{Division of Science, NYU Abu Dhabi, PO Box 129188, Abu Dhabi, UAE}
\affil{Center for Astrophysics and Space Science (CASS), NYU Abu Dhabi, PO Box 129188, Abu Dhabi, UAE}
\affil{Yunnan Observatory, Kunming, Yunnan, CN}
\author[0000-0003-4679-1058]{J. D. Gelfand}
\affil{Division of Science, NYU Abu Dhabi, PO Box 129188, Abu Dhabi, UAE}
\affil{Center for Astrophysics and Space Science (CASS), NYU Abu Dhabi, PO Box 129188, Abu Dhabi, UAE}

\begin{abstract}
Synchrotron X-ray emission from a pulsar wind nebula (PWN) is a sensitive probe of its magnetic field and high energy particle population.
Here we analyze contemporaneous \nustar \ and \xmm \ observations of the PWN \src, powered by pulsar \srcp.
We also present a preliminary timing analysis the central pulsar \srcp, and analyze its X-ray pulse profiles in different energy bands.
We detect X-ray emission from the combined pulsar and PWN system up to $\approx70$ keV, while emission from the PWN itself has been detected up to $\approx30$~keV, with a photon index $\Gamma$ increasing from $\sim 1.9$ to $\sim 2.4$ with photon energy between $3-30$ keV.
PWN \src's \ X-ray spectrum is consistent with a broken power law, with break energy $E_{\rm break} \approx 5$~keV, consistent with synchrotron cooling of a single powerlaw particle spectrum. 
The best fit broadband SED model after the inclusion of this new spectral data indicates a maximum particle energy  $E_{\rm max}\sim$~400~TeV.
We discuss \srcp \ and PWN \src \ in the context of other PWNe powered by young energetic pulsars.
\end{abstract}

\section{Introduction}
\label{introduction}

Young rotation-powered pulsars are known sources of intense non-thermal X-ray emission.
These non-thermal X-rays can be produced by relativistic particles in both the neutron star (NS) magnetosphere and surrounding pulsar wind nebula (PWN).
In the case of PWN emission, the $e^{\pm}$ streaming through the PWN  emit both synchrotron and inverse Compton (IC) radiation, which can be observed across the electromagnetic spectrum,  from radio to $\gamma$-rays (see \cite{Slane2017} for a review of PWNe).
At X-ray energies, the dominant contribution is synchrotron emission, whose properties depend on both the magnetic field strength and the particle spectrum within the PWN.
In contrast, the PWN's IC emission, which dominates the SED at $\gamma-$ray energies, is determined by the particle and target photon spectra and is independent of the magnetic field.
Therefore modeling the broadband PWN SED probes both the $B$ field and particle spectrum within the PWN.

PWN \src \ is similar in many respects to the Crab Nebula, with Chandra observations revealing that both are young PWNe  characterized by a central pulsar point source, a ring, jets, and a surrounding diffuse synchrotron nebula \citep{Lu2002, Temim2010, Bocchino2010}.
\src \ is among the several PWNe (e.g. 3C 58 and the Crab Nebula) that lack an obvious shell associated with a supernova remnant (SNR) forward shock.
It is not clear if the SNR ejecta have not yet been shocked, or more sensitive X-ray and radio observations will reveal the SNR shell. 
Radio and X-ray shells associated with \src's forward shock have been suggested, but not confirmed \citep{Goedhart2024, Tsalapatas2024, Bocchino2010}.

\src \ is powered by the 137~ms pulsar \srcp, with a large spin-down power $\dot{E} \equiv 4 \pi^2 I \frac{P}{\dot{P}^3} = 1.2 \times 10^{37}$ erg s$^{-1}$ and young characteristic age $\tau_{\rm ch} \equiv P/2\dot{P} \approx 2900$ yr \citep{Camilo2002}.
The pulsar X-ray spectrum is non-thermal and is consistent with a single powerlaw with a photon index $\Gamma = 1.44 \pm 0.04$ \citep{Temim2010}.
Similarly, the PWN's X-ray spectrum is also non-thermal, with a photon index $\Gamma \sim 1.8$ and an unabsorbed flux $\sim 5 \times$ greater than the pulsar the 0.2$-$10 keV band.

Recently, the Large High Altitude Air Shower Observatory (LHASSO) reported the detection of $\gtrsim 100$ TeV $\gamma$-rays from 43 sources \citep{Cao2024}.
\src \ is located $\leq 0.29^{\circ}$ from one of these $\gamma$-ray sources:  \lhaasosrc.
\src \ may therefore be a PeVatron like the Crab Nebula, although there are other potential $\gamma$-rays sources nearby.

Measurements of \src's X-ray synchrotron emission are required to investigate \src's potential to accelerate particles to energies $\gtrsim 100$ TeV.
We have obtained coordinated \nustar \  and \xmm \ observations of \src, in order measure its  broadband (0.5 – 80 keV) X-ray spectrum.
These X-ray spectral measurements are also essential inputs in broadband SED modeling of \src, which  will also help determine if \src \ is producing the $\gtrsim 100$ TeV $\gamma$-rays  observed by LHAASO.

In this paper, we present a comprehensive analysis of the broadband X-ray spectrum of the pulsar \srcp  \ and its PWN G54.1$+$0.3 with these \xmm \ and \nustar \ observations. 
This paper is structured as follows: 
In Section 2, we describe the observations, data reduction and analysis methods.
In Section 3, we present the results of the temporal and spectral analysis of these data. 
In Section 4, we present a SED model of the PWN in \src, updated with these new X-ray spectral data points.
We discuss the implications of these results in Section 5, and summarize our findings in Section 6.

\section{Data reduction}
\label{data_reduction}

\begin{deluxetable*}{llccc}
\label{tab:xray_obs}
\tablecaption{Log of X-ray Observations}
\tablehead{\colhead{Date} & \colhead{Observatory}  & \colhead{ObID} & \colhead{Instr./Mode} & \colhead{Exposure}   \\
\colhead{(UT)} &  & & & \colhead{(ks)} }
\startdata
2016 Mar 27 & \xmm & 0762980101 & EPIC-MOS1/Full Frame  & 108.7 \\
... & ... & ... & EPIC-MOS2/Full Frame  & 108.6 \\
... & ... & ... & EPIC-pn/Full Frame & 108.9 \\
\hline
2016 Mar 27 & \nustar & 40101006002 & FPMA & 54.3  \\
... & ... & ... & FPMB &  54.1  \\
\hline
2016 Jul 2  & \nustar & 40201012002 & FPMA & 80.1 \\
... & ... & ... & FPMB & 80.0  \\
\enddata
\end{deluxetable*}

\subsection{\nustar\ }

\begin{figure*}
\centering
\includegraphics[width=1.0\linewidth]{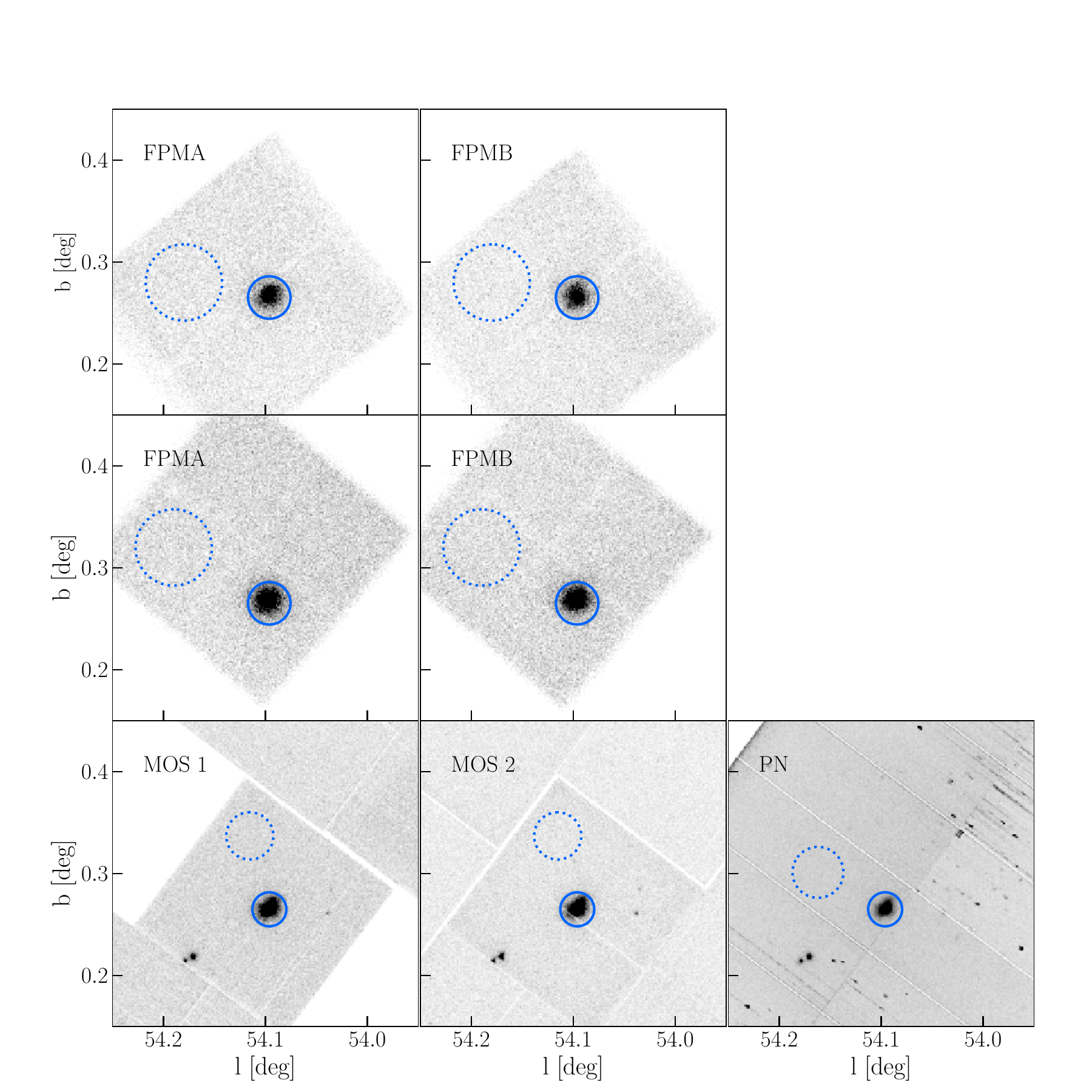}
\caption{X-ray images of \src. Solid (dotted) circles indicate the source (background) spectral extraction regions for the \nustar \ and \xmm \ observations.}
\label{fig:source_background_regions}
\end{figure*}

\nustar \ \citep{Harrison2013} observed \src \ on 2016 March 27  and again on 2016 July 02 (Table \ref{tab:xray_obs}). 
The standard data reduction pipeline {\sc nupipline} was run with CALDB version 20240701.
Spectra were extracted from the two \nustar \ focal plane modules, FPMA and FPMB, using  circular regions with a 1$\arcmin$\!.25 radius centered at the source position, and circular background regions with a 2$\arcmin$\!.25 radius (see Figure \ref{fig:source_background_regions}).
The spectra were binned to ensure a minimum of 25 counts per bin.

\begin{figure*}
\centering
\includegraphics[width=01.04\linewidth]{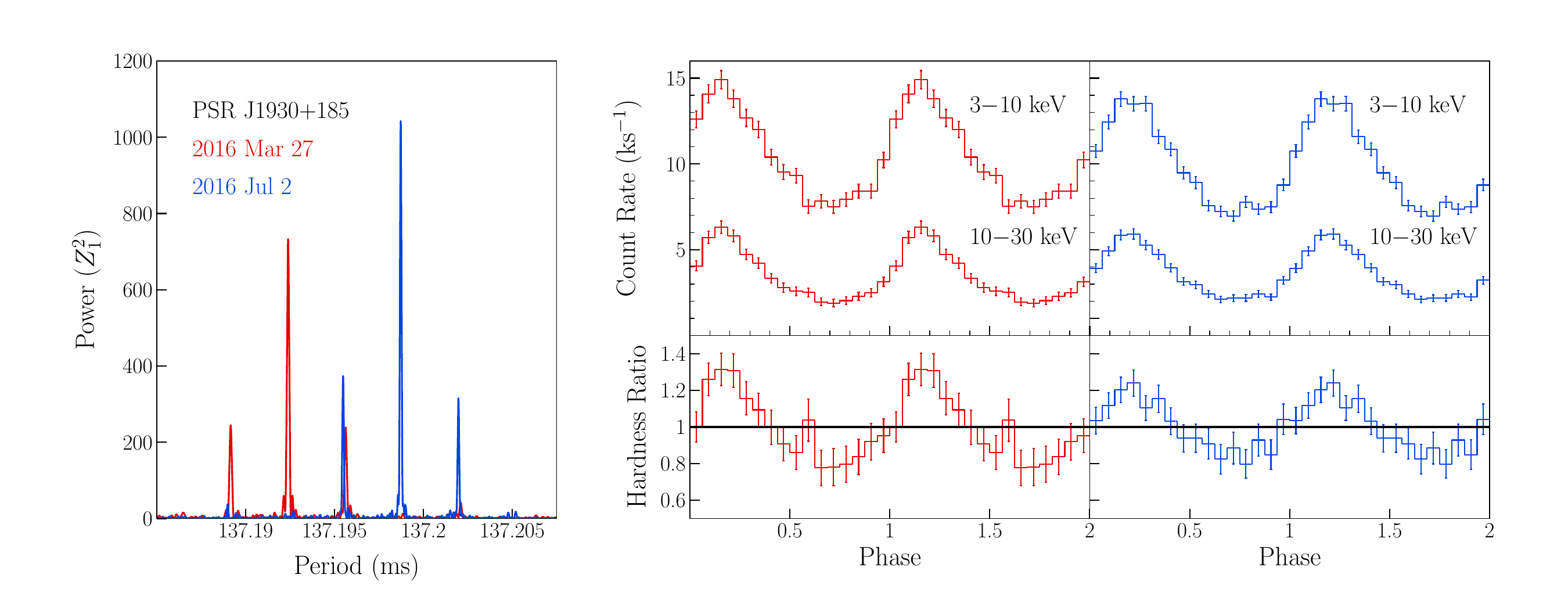}
\caption{ Left: $Z_1^2$ power density spectra in each of the two \nustar \ observations, with the central peaks at each epoch corresponding to periods $P = 137.192387(3)$~ms and $P = 137.198727(2)$~ms. 
Upper Right: Background subtracted pulse profiles in the 3$-$10~keV and 10$-$30~keV bands. 
Lower Right: Normalized hardness ratio (10$-$30~keV count rate / 3$-$10~keV count rate) as a function of rotational phase.   }
\label{fig:timing}
\end{figure*}

\subsection{\xmm }

\xmm\ observed \src \ for 108~ks \ on 2016 March 27 (ObsID 0762980101).
The \xmm\ data was analyzed using  the  standard XMM scientific  
analysis  software (SAS) version {\sc xmmsas\_20160201\_1833-15.0.0} with the latest calibration files. 
We created reprocessed event files for the EPIC-PN,  EPIC-MOS1 and EPIC-MOS2 using the tasks {\sc epproc} and {\sc emproc}, respectively.

Time intervals affected by solar flares were removed, reducing the useful exposure time to $\sim$52 ks.
We then extracted source spectra from PN, MOS1 and MOS2 using circular regions with a 1$\arcmin$ radius radius centered at the source position.
We selected circular source-free background regions with a  1$\arcmin$\!.5 radii on the same CCD (Figure \ref{fig:source_background_regions}). 
We created redistribution matrices and ancillary files using the SAS tools {\sc rmfgen} and  {\sc arfgen}, respectively.
The spectra were binned to ensure a minimum of 25 counts per bin.

\subsubsection{Search for Diffuse X-ray Emission}
\label{diffuse}

\begin{figure*}
\centering
\includegraphics[width=1.0\linewidth]{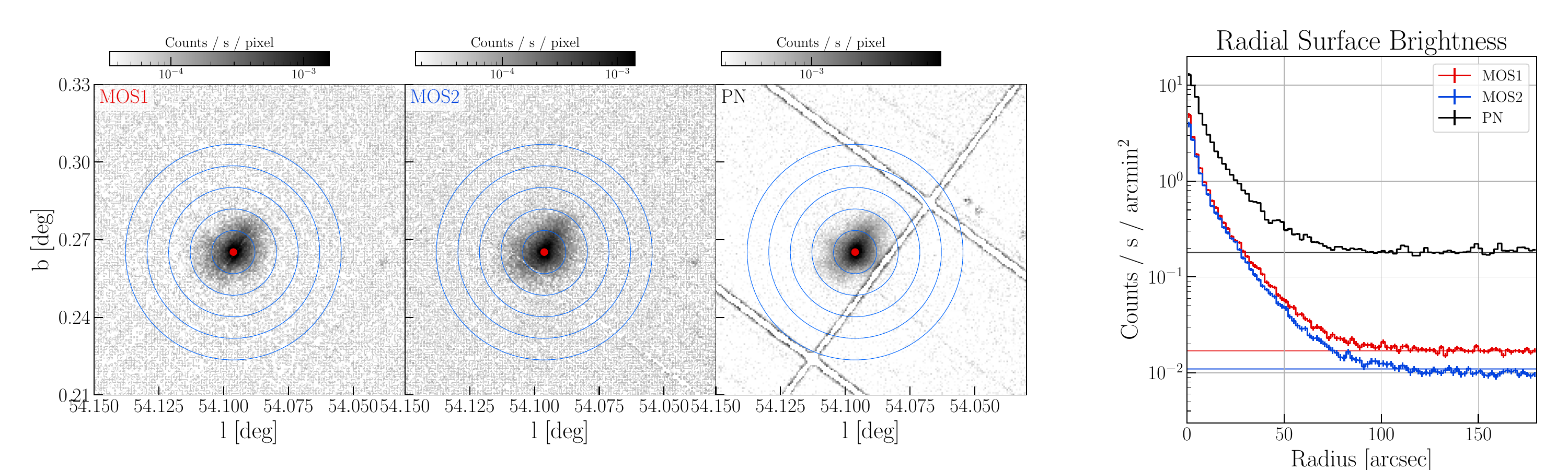}
\caption{Left: Exposure and vignetting corrected  \xmm \ images of \src, scaled to highlight weak diffuse X-rays from the PWN.  The brightest emission coincides with the position of the pulsar (red point), and blue rings indicate radial distance in units of 30 arcseconds. Right: Radial surface brightness measured by the \xmm \ MOS1, MOS2 and PN detectors. Horizontal lines indicate the background surface brightness of each detector.}
\label{fig:xmm_diffuse}
\end{figure*}

Previous studies reported diffuse X-ray emission extending out to $\approx 400$ arcsec from \srcp, potentially associated with \src's SNR shell \citep{Bocchino2010}. 
Figure \ref{fig:xmm_diffuse} shows zoomed-in, exposure and vignetting corrected images from the \xmm \ MOS1, MOS2 and PN detectors. 
The radial surface brightness profiles in the right panel of Figure \ref{fig:xmm_diffuse} indicate the diffuse emission extends out no more than $100^{"}$ from \srcp. 
This X-ray extension of the \src \ PWN is consistent with the $\sim 80^{"}$ radius of the infrared dust shell \citep{Temim2017}.

\section{Data Analysis}
\subsection{Timing Analysis}
\label{timing}

We searched for \srcp's pulsed X-ray emission by extracting photons from within a 1 arcminute radius around the pulsar, and calculating the photon arrival times at the Solar System barycenter.
The photon arrival times from the FPMA and FPMB detectors were combined to calculate the power density spectra shown in Figure \ref{fig:timing} \citep{Leahy1983}. 
We identified a clear periodic signal at $\approx$ 137 ms (Figure \ref{fig:timing}), consistent with previous detections \citep{Camilo2002,Lu2007}. 
The central peaks at each epoch correspond to pulsar rotation periods $P = 137.192387(3)$~ms and $P = 137.198727(2)$~ms.
Table \ref{tab:NS_period_ob} lists these two \nustar\ measurements of PSR J1930$+$1852's rotation period, along with previous measurements. 

\cite{Camilo2002} used radio observations separated by 8 months to measure $\dot{P} = 7.5057(1) \times 10^{-13}$~s~s$^{-1}$.
Extrapolating from their precisely measured 2002 Jan 17 spin period to the \nustar \ measured periods in Table \ref{tab:NS_period_ob}, we calculate a larger long term $\dot{P} = 7.517(3) \times 10^{-13}$~s~s$^{-1}$, consistent with the $\dot{P} = 7.514(4) \times 10^{-13}$~s~s$^{-1}$ inferred from the change in spin period between the two \nustar \ observations.
This increase in the long term $\dot{P}$ implies that \srcp's braking index $n < 2$. 
A detailed timing analysis will be discussed by Alford et al. 2025 (in preparation).

In order to check for phase-dependent spectral variations, we calculated the hardness ratio (HR) in each phase bin. 
We define HR$=N_{h}/N_{s}$, where $N_{s}$ and $N_{h}$ are the count rates in the $3-10$ and $10-30$ keV energy bands, respectively. 
Normalized hardness ratios are shown in the lower right panels of Figure \ref{fig:timing}. 
We find that \srcp's on-pulse emission is harder than the off-pulse emission.

\begin{deluxetable*}{lcll}
\label{tab:NS_period_ob}
\tablecaption{Measurements of \srcp's rotation period.}
\tablehead{\colhead{Date} & \colhead{Observatory} & \colhead{Period}  & \colhead{Reference}  \\
\colhead{(UT)} & & \colhead{(ms)}  &  }
\startdata
1997 Apr 27 & {\it ASCA} & 136.74374(5)   & 1 \\
2002 Jan 17 & {\it Arecibo}  & 136.855046957(9)  & 1 \\
2002 Sep 12 & {\it RXTE}  & 136.871312(4) & 2 \\
2002 Dec 23 & {\it RXTE}  & 136.877919(3) & 2 \\
2003 Jun 30 & {\it Chandra} & 136.890130(5) & 2 \\
2016 Mar 27 & {\it NuSTAR}  & 137.192387(3) & This work \\
2016 Jul 2  & {\it NuSTAR}  & 137.198727(2) & This work \\
\enddata
\tablerefs{
(1) \citet{Camilo2002};
(2) \citet{Lu2007};
}
\end{deluxetable*}

\subsection{Spectral Analysis}
\label{spectral}

\begin{figure*}
\centering
\includegraphics[width=1.0\linewidth]{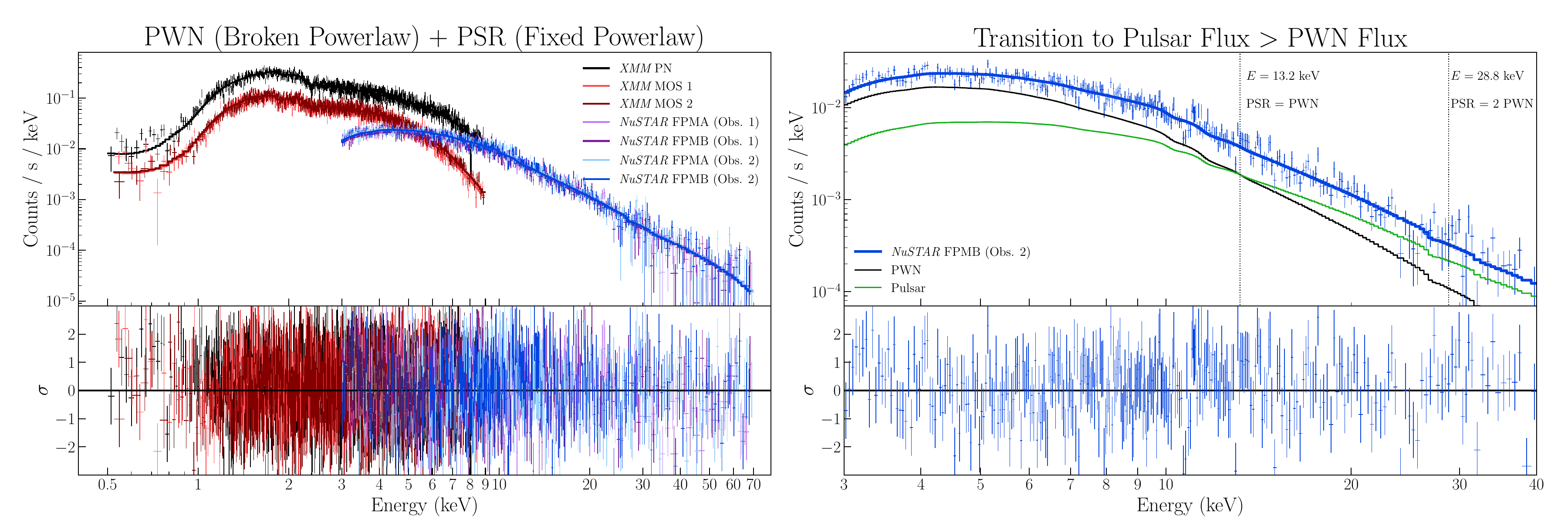}
\caption{Left: Joint \xmm\ and \nustar\ spectra fit with a broken power law plus pulsar power law ({\tt BKN + POW}) model. The parameters of the pulsar powerlaw model were fixed to the values determined from a previous Chandra observation.  Right: The \src \ spectrum from the second, longer   \nustar \ observation, focusing on the energy range where the pulsar flux becomes dominant over the PWN flux. 
}
\label{fig_pha_xmm_nustar_all}
\end{figure*}

X-ray spectral analysis was performed using {\tt HEASoft} version 16.33 and of the {\tt Xspec} version 12.14.0 \citep{Arnaud1996}.
Throughout this spectral modeling, we have used the Tuebingen–Boulder X-ray absorption model {\tt tbabs} with elemental abundances from \cite{Wilms2000}.

The \nustar \ spectrum of the G54.1 PWN includes X-rays from the pulsar \srcp, because \nustar \ cannot spatially resolve the pulsar from the PWN.
Since we want to include only PWN emission in our SED modeling, we include a powerlaw spectral component in our analysis to account for the spectrum of \srcp.
We  explored a phase-resolved spectroscopic analysis, treating the off-pulse spectrum as pure background PWN emission, and found that the derived pulsar spectrum differed from the uncontaminated Chandra spectrum reported by \cite{Temim2010}.  
This could be due to significant unpulsed non-thermal X-ray emission from the pulsar contaminating the "off" spectrum.   
Considering the large NuSTAR PSF, we chose to adopt the results of the Chandra analysis for the pulsar spectrum.
The powerlaw index $\Gamma$ and normalization of the pulsar component were fixed to the values obtained from a previous \chan\ observation with $\Gamma_{\rm psr}=1.44\pm0.04$ \citep{Temim2010}.
We have confirmed that \cite{Temim2010} used the same Wilms abundances in their absorption model (Temim, private communication).

Holding the pulsar's spectral parameters fixed, we then fit models for the pulsar and PWN emission to the joint \xmm-\nustar\ spectra.
We include data from the two XMM MOS detectors, the XMM PN detector, and the two \nustar \ detectors.
We explored a power law model and a broken power law model ({\sc const*tbabs(pow$+$bknpow)}, hereafter PL$+$BKN) or a two single power law models ( {\sc const*tbabs(pow$+$pow)}, hereafter PL$+$PL) in the 0.5 $-$ 70 keV energy band. 
The constant factors {\sc const} were multiplied by the spectra from each detector in order to account for systematic calibration offsets.
The factors differed by $< 1\%$ between the \nustar \ A and B modules, and $< 10\%$ between \xmm \ and \nustar.

The joint \xmm-\nustar \ spectrum (Figure \ref{fig_pha_xmm_nustar_all}) shows curvature, especially above $\sim5$~keV, which is more consistent with the PL$+$BKN model than the PL$+$PL model, and this is reflected in the $\chi^2$ values listed in Table \ref{tbl:spectral_params}.
We checked that the spectral curvature is intrinsic to the \src \ spectrum, and not an artifact of the relative \xmm-\nustar calibration, by confirming that the \xmm \ data alone are more consistent with the PL$+$BKN model versus the PL$+$PL model.
We explored keeping the absorption column fixed at the value $N_{\rm H} =(1.95\pm0.03)\times10^{22}~{\rm cm}^{-2}$ found by \cite{Temim2010} and also allowing $N_{\rm H}$ to vary.
A slightly lower $N_{\rm H}\approx 1.5\times10^{22}~{\rm cm}^{-2}$ values is preferred by the joint \xmm-\nustar\ spectra.
Table \ref{tbl:spectral_params} lists the spectral modeling results for both the PL$+$BKN and PL$+$PL models, for both $N_{\rm H}$ values.
The left panel of Figure \ref{fig_pha_xmm_nustar_all} shows the joint fit of the \nustar \ and \xmm \ data to the PL$+$BKN model.
The right panel of Figure \ref{fig_pha_xmm_nustar_all} shows the shows the individual contributions of the pulsar and PWN, and indicates that the PWN contributes substantially 
up to $\sim30$~keV.

Since we found that the joint \xmm-\nustar\ spectra contains significant curvature, we also fit the spectrum in the 3$-$6 and 6$-$30 keV bands individually.
We did this while allowing the $N_{\rm H}$ values to vary, and we fixed the $N_{\rm H}$ values to the corresponding values in Table \ref{tbl:spectral_params}.
These results are shown in Table \ref{tab:main_results}, and are the observational data points that we use in our SED modeling in Section \ref{section:SED}.

A powerlaw distribution $\frac{dN}{dE}$ of synchrotron cooling $e^{\pm}$ with an initial constant particle spectral index $p$ will develop a particle spectrum break $\Delta p = 1$, corresponding to an observed photon spectral break $\Delta \Gamma = 0.5$ \citep{kar1962,ryb1979,rey2009}. 
The $\Delta \Gamma \approx 0.5$ spectral break at $E_{\rm break}\approx5$~keV is therefore consistent with the synchrotron cooling. 
A synchrotron cooling break will be observed in a system of synchrotron emitting particles at a photon energy $E_{\rm break}$ if the corresponding particle energy $E$ at the synchrotron spectrum peak:
\begin{equation}
    E_{\rm break} = h \nu_{\rm peak} = 0.29 \times \frac{3}{2} \left( \frac{E}{m_{e} c^2}\right)^2  \left(\frac{e B}{m_e c} \right),
\end{equation}
is equal to the system age $\tau$.
If we know $\tau$, then the value of $E_{\rm break}$ can then be used to estimate the system's magnetic field strength:
\begin{equation}
     E_{\rm break} = 168  \left(\frac{B}{5~\mu~\rm{G}}\right)^{-2} \left(\frac{\tau}{3~ \rm{kyr}}\right)^{-1}    \mathrm{TeV}
\end{equation}
The SED modeling presented in Section 4 makes an independent estimate the PWN age $\tau$ and magnetic field $B$, and therefore provides an independent estimate of $E_{\rm break}$.

\begin{deluxetable*}{ccccccccc}
\tablecaption{Joint Fits to \xmm\ and \nustar\ Spectra.}
\tablehead{
\colhead{Model} &  \colhead{$N_{\rm H}$} & \colhead{$\Gamma_{\rm psr}$} & \colhead{Pulsar Flux$^a$} & \colhead{$\Gamma_{1}$} & \colhead{$E_{\rm break}$}  & \colhead{$\Gamma_{2}$}  & \colhead{PWN Flux$^a$} & \colhead{$\chi^2$ (d.o.f.)}  \\
 &  \colhead{($10^{22}~\mathrm{cm}^{-2}$)} & & \colhead{($10^{-12}$~erg~cm$^{-2}$~s$^{-1}$)} & &  \colhead{(keV)} & & \colhead{($10^{-12}$~erg~cm$^{-2}$~s$^{-1}$)}}
\startdata
PL$+$PL &  $1.73_{-0.03}^{+0.02}$  &  1.4 (fixed) & 5.1 (fixed) &   $2.10_{-0.03}^{+0.03}$ & ...  &  ...  & $6.6_{-0.1}^{+0.1}$ & 3881 (3736) \\ 
PL$+$PL& $1.71_{-0.03}^{+0.02}$  &  1.44 (fixed) & 5.1 (fixed) &  $2.06_{-0.02}^{+0.02}$ & ...   & ...  & $6.4_{-0.1}^{+0.1}$ & 3822 (3736) \\ 
PL$+$PL&  $1.69_{-0.02}^{+0.02}$  &  1.48 (fixed) & 5.1 (fixed) &  $2.02_{-0.02}^{+0.02}$ & ...  & ...  & $6.4_{-0.1}^{+0.1}$ & 3781 (3736)  \\
\hline
PL$+$BKN& $1.51_{-0.04}^{+0.03}$  &  1.4 (fixed) & 5.1 (fixed) &  $1.83_{-0.06}^{+0.04}$ & $5.5_{-0.5}^{+0.4}$ & $2.46_{-0.08}^{+0.07}$ & $5.8_{-0.1}^{+0.1}$ & 3618 (3734)  \\
PL$+$BKN& $1.50_{-0.03}^{+0.04}$  &  1.44 (fixed) & 5.1 (fixed) &  $1.78_{-0.08}^{+0.08}$ & $5.1_{-0.3}^{+0.6}$ & $2.32_{-0.05}^{+0.07}$ & $5.8_{-0.1}^{+0.2}$ & 3572 (3734)  \\
PL$+$BKN& $1.49_{-0.03}^{+0.03}$  &  1.48 (fixed) & 5.1 (fixed) &  $1.75_{-0.05}^{+0.04}$ & $5.0_{-0.4}^{+0.4}$ & $2.24_{-0.05}^{+0.05}$ & $5.8_{-0.1}^{+0.1}$ & 3564 (3734)   
\enddata
\tablecomments{The pulsar contribution to the combined pulsar plus PWN spectrum is modeled with a single power law, with its flux and photon index $\Gamma_{\rm psr}$ held fixed at the {\it Chandra} derived values \citep{Temim2010}.
}
\label{tbl:spectral_params}
\tablenotetext{a}{Unabsorbed Flux in the 3$-$30 keV band}
\end{deluxetable*}

\begin{deluxetable*}{lccccc}
\tablecaption{G54.1$+$0.3 PWN Spectrum in the 3$-$30 keV Band}
\tablehead{ \colhead{Energy Band} & \colhead{$N_{\rm H}$} & \colhead{$\Gamma_{\mathrm{psr}}$} & \colhead{Pulsar Flux$^a$}  & \colhead{$\Gamma_{\mathrm{pwn}}$} & \colhead{PWN Flux$^a$} \\
 \colhead{(keV)}  & \colhead{($10^{22}~\mathrm{cm}^{-2}$)} & & \colhead{($10^{-12}$~erg~cm$^{-2}$~s$^{-1}$)} & & \colhead{($10^{-12}$~erg~cm$^{-2}$~s$^{-1}$)} }
\startdata
\hline
3$-$6    & 1.51 (fixed)  &  1.4 (fixed) & 0.9 (fixed) &  $1.95_{-0.06}^{+0.06}$ & $2.30_{-0.04}^{+0.04}$   \\
6$-$30  & 1.51 (fixed)  &  1.4 (fixed) & 4.2 (fixed) & $2.45_{-0.07}^{+0.07}$ & $3.89_{-0.15}^{+0.11}$    \\
\hline
3$-$6    & 1.50 (fixed)  &  1.44 (fixed) & 0.9 (fixed) & $1.91_{-0.06}^{+0.06}$ & $2.30_{-0.11}^{+0.09}$  \\    
6$-$30  & 1.50 (fixed)  &  1.44 (fixed) & 4.2 (fixed) & $2.33_{-0.06}^{+0.06}$ & $3.83_{-0.11}^{+0.09}$  \\
\hline
3$-$6    & 1.49 (fixed)  &  1.48 (fixed) & 0.9 (fixed) & $1.90_{-0.05}^{+0.05}$ & $2.39_{-0.14}^{+0.10}$   \\
6$-$30  & 1.49 (fixed)  &  1.48 (fixed) & 4.2 (fixed) & $2.24_{-0.06}^{+0.06}$ & $3.89_{-0.13}^{+0.10}$   \\
\enddata
\tablenotetext{a}{Unabsorbed PWN flux in the corresponding energy bands}
\label{tab:main_results}
\end{deluxetable*}

\section{SED Modeling}
\label{section:SED}

With the benefit of hard X-ray constraints from \nustar, we can extend the G54.1$+$0.3 SED dynamical modeling described in \cite{Gelfand2015}.
This is a one-zone, energy-conserving radiative model, that tracks the time evolution of the G54.1$+$0.3 system \cite{Gelfand2009}.
This model predicts the broadband PWN spectral energy distribution (SED), the sizes on the SNR and PWN, and is constrained by the presently measured values of the pulsar period $P$ and period derivative $\dot{P}$ \citep{Hattori2020,Straal2023,abd2023,Pope2024}.

Table \ref{tab:model_obs} lists the observed properties of the G54.1$+$0.3 system along with the values predicted by the best fit model.
The uncertainty in the pulsar X-ray flux and photon index contributes a systematic uncertainty to the G54.1$+$0.3 PWN X-ray spectrum.
For the purposes of this modeling, we adopt the range X-ray fluxes and photon indices listed in Table \ref{tab:main_results}, corresponding to a range in values of the pulsar spectrum photon index (1.40$-$1.48).
Following the methodology of \cite{Hattori2020}, we consider all values of the X-ray fluxes within this range equally consistent with the model, and we accordingly only add values outside of this range when calculating the $\chi^{2}$ values.

\begin{figure*}
    \centering               
    \includegraphics[width=.95\linewidth]{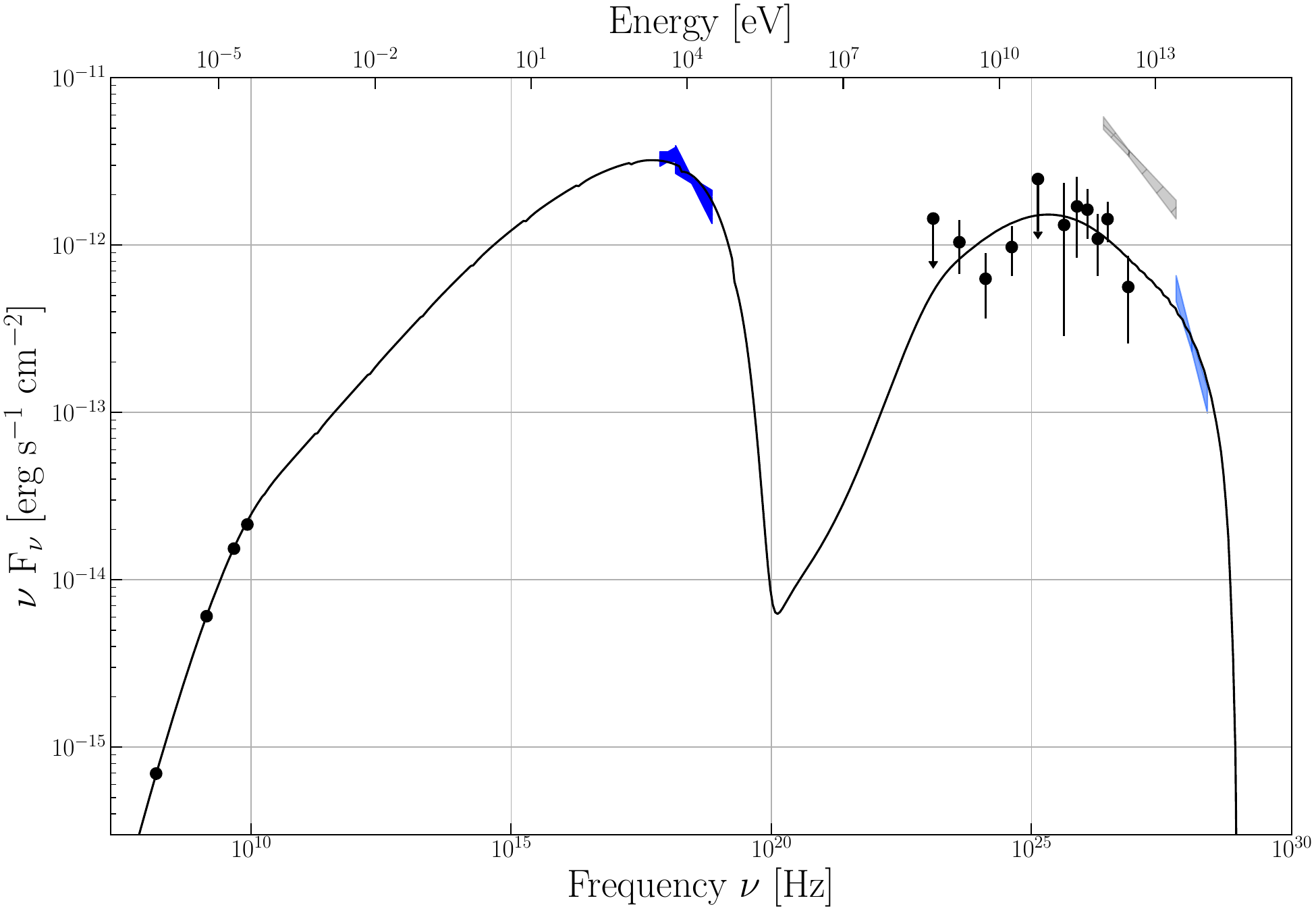}
    \caption{A PWN dynamical evolution model fit to broadband SED (and radius) of the PWN in G54.1$+$0.3. The model is described in Section \ref{section:SED}, the input model parameters are given in Table \ref{tab:model_pars}, and the model predictions are compared to the observed properties of the G54.1$+$0.3 system in Table \ref{tab:model_obs}.}
    \label{fig:SED_model}
\end{figure*}

In addition to the original data points described in \cite{Gelfand2015}, we also include a 150 MHz radio data radio flux and several gamma-ray photon densities in this new analysis.
This low frequency 150 MHz radio data point was obtained from the LOFAR data archive \citep{Heald2015}.
The new Fermi gamma-ray data points were obtained from \cite{Eagle2022}.
We also include some of the LHAASO gamma-ray data reported by \cite{Cao2024}.
The 1$-$25 TeV LHAASO flux is significantly higher than the VERITAS data \citep{Acciari2010}, likely due to the contribution of one or more unrelated sources, and we plot this flux in gray in Figure \ref{fig:SED_model} for reference, but do not fit to it in our PWN modeling.
We do include the 25$-$100 TeV LHAASO flux in our modeling.
The 25$-$100 TeV  LHAASO flux is reported as the best fit to an assumed powerlaw, while our PWN model predicts more detailed spectral curvature in this band.

Table \ref{tab:model_pars} lists the best fit model input parameters, all of which were varied to fit the PWN SED and angular size.
The $\chi^2 = 28.9$ (8 d.o.f.) statistic is dominated by the deviation of the photon index $\Gamma$ from the LHAASO photon index.
This is expected because the PWN model predicts spectral curvature in this band while at this time only a powerlaw fit has yet been reported by LHAASO.
The observed 25$-$100 TeV LHAASO flux is in excellent agreement with the model predicted value.
In Figure \ref{fig:SED_model} we have plotted the SED predicted by the best fit PWN dynamical evolution model.

\begin{table*} 
    \caption{PWN Model Input Parameters}
    \label{tab:model_pars}
    \centering
    \begin{tabular}{lc}
    \hline
    \hline 
    {\sc Parameter} & {\sc Value}   \\
    \hline
    \multicolumn{2}{c}{\it SNR Parameters} \\ 
    Explosion Energy $E_{\rm sn}$ & $8.9\times10^{50}~{\rm erg}$  \\
    Ejecta Mass $M_{\rm ej}$ & 20.9~M$_\odot$  \\
    ISM Density $n_{\rm ism}$ & $8.5\times 10^{-3}~{\rm cm}^{-3}$  \\
    Distance $D$ &  6.2~kpc \\
\hline
     \multicolumn{2}{c}{\it PWN Parameters} \\ 
    Wind Magnetization $\eta_{\rm B}$ & $2.4 \times 10^{-3}$  \\
    Minimum e$^{\pm}$ Injection Energy $E_{\rm min}$ & 10~GeV  \\
    Maximum e$^{\pm}$ Injection Energy $E_{\rm max}$ & 400~TeV  \\
    Particle Index $p_{1}$ & 2.29  \\
    Particle Index $p_{2}$ & 2.42  \\
    External Photon Field Temperature  $T_{\rm ic}$ & 2,750~K  \\
    External Photon Field Normalization  $K_{\rm ic} a T_{\rm ic}^{4}$ & $25.3~{\rm eV~cm^{-3}}$  \\
\hline
    \multicolumn{2}{c}{\it Pulsar Parameters} \\   
    Spin-down Timescale $\tau_{\rm sd}$ & 3.57~kyr \\
    Braking Index $p$  & 1.90 \\

    \hline
    \hline
    \end{tabular}
\end{table*}

\begin{table*}[ht!]
\caption{Observed properties of the G54.1$+$0.3 system, alongside the model predicted properties}
    \label{tab:model_obs}
    \resizebox{0.95\linewidth}{!}{
    \centering
    \begin{tabular}{lccc} 
    \hline
    \hline 
    {\sc Property} & {\sc Observed} & {\sc Model} & {\sc References} \\
    \hline
    \multicolumn{4}{c}{\it Pulsar Properties} \\
    \hline
    $\dot{E}$ & $1.2\times10^{37} \rm{erg \ s^{-1}}$ & Fixed & 1\\
    $\tau_{\rm char}$ & 2900 yr & Fixed & 1 \\
    \hline
    \multicolumn{4}{c}{\it Pulsar Wind Nebula Properties} \\
    \hline
    PWN Radius $\theta_{\rm pwn}$ & $1\farcm14 \pm 0\farcm04$ & $1\farcm13$ & 2,3 \\
    0.15 GHz Flux Density & $464\pm47$ mJy & 459 &  4 \\
    1.4 GHz Flux Density  & $433\pm30$ mJy & 432 &  2 \\
    4.7 GHz Flux Density  & $327\pm25$ mJy & 331 &  2 \\
    8.5 GHz Flux Density  & $252\pm20$ mJy & 261 &  2 \\
    Flux (3$-$6~keV)$^a$ & 2.30$_{-0.11}$ $-$2.39$^{+0.11}$ & $2.16$ & This work\\
    Photon Index $\Gamma$(3$-$6~keV) & 1.90$_{-0.05}$ $-$ 1.95$^{+0.06}$ & $2.04$ & This work \\
    Flux (6$-$30~keV)$^a$ & 3.83$_{-0.11}$ $-$ $3.89^{+0.11}$ & $3.90$ & This work\\
    Photon Index $\Gamma$(6$-$30~keV) & 2.24$_{-0.06}$ $-$ 2.45$^{+0.07}$ & $2.56$ & This work \\  
    535 MeV Photon Density$^b$ & $<3.14\times10^{-6}$ & $1.10 \times10^{-6}$ & 5 \\
    1.7 GeV Photon Density$^b$ & $(2.23 \pm 0.80) \times10^{-7}$ & $1.62 \times10^{-7}$ & 5 \\
    5.4 GeV Photon Density$^b$ & $(1.33 \pm 0.56)  \times10^{-8}$ & $2.09 \times10^{-8}$ & 5 \\
    17.3 GeV Photon Density$^b$ & $(2.03 \pm 0.67) \times10^{-9}$ & $2.49 \times10^{-9}$ & 5 \\
    55.1 GeV Photon Density$^b$ & $<5.10\times10^{-10}$ & $2.68 \times10^{-10}$ & 5 \\
    176 GeV Photon Density$^b$ & $(2.67 \pm 2.09) \times10^{-11}$ & $2.53 \times10^{-11}$ & 5 \\
    311 GeV Photon Density$^b$ & $(1.10 \pm 0.56)  \times10^{-11}$ & $7.92 \times10^{-12}$ & 6 \\
    492 GeV Photon Density$^b$ & $(4.2 \pm 1.4) \times10^{-12}$ & $3.06 \times10^{-12}$ & 6 \\
    780 GeV Photon Density$^b$ & $(1.12 \pm 0.45) \times10^{-12}$ & $1.02 \times10^{-12}$ & 6 \\
    1.2 TeV Photon Density$^b$ & $(6.2 \pm 1.7) \times10^{-13}$ & $3.90 \times10^{-13}$ & 6 \\
    3.0 TeV Photon Density$^b$ & $(3.1 \pm 2.1) \times10^{-14}$ & $5.2 \times10^{-14}$ & 6 \\ 
    25$-$100~TeV Photon Index $\Gamma$  & $3.11\pm0.12$ & 2.69 & 7  \\
    25$-$100~TeV Differential Flux$^d$ $N_0$  & $0.64\pm0.06$  & 0.66 & 7  \\
    \hline
    \hline
    $\chi^2$ (d.o.f.) &  & 28.9 (8)  \\
    \hline
    \end{tabular}
}

\tablenotetext{a}{Unabsorbed flux in units of $10^{-12}$ ergs s$^{-1}$ cm$^{-2}$.}
\tablenotetext{b}{Photon density in units of photons s$^{-1}$ cm$^{-2}$ TeV$^{-1}$.}
\tablenotetext{c}{$\frac{\mathrm{d}N}{\mathrm{d}E}  = N_0 \left(\frac{E}{50~{\rm TeV}}\right)^{-\Gamma}$ with  $N_0$ in units of $10^{-16}$ photons s$^{-1}$ cm$^{-2}$ TeV$^{-1}$}
\tablerefs{
(1) \citet{Camilo2002};
(2) \citet{Lang2010};
(3) \citet{Lu2002};
(4) \citet{Heald2015};
(5) \citet{Eagle2022};
(6) \citet{Acciari2010};
(7) \citet{Cao2024}
}
\end{table*}

\section{Discussion}
\label{discussion}

\subsection{Comparison with Previous Work}

\cite{Gelfand2015} modeled G54.1+0.3's SED, without the additional hard X-ray, low frequency radio and $\gamma$-ray data considered in this analysis. 
Also, \cite{Gelfand2015} fit the PWN model to a SNR size, which we have not done in this analysis, since the SNR detection is now uncertain. 

The \nustar \ data analysis presented in this paper has allowed us to constrain the maximum particle energy in the PWN. 
The best fit model predicts a maximum particle energy of $\approx0.4$~PeV, which is significantly less than the best fit value 0.96$-$2700 PeV range that \cite{Gelfand2015} calculated without the constraints from this \nustar \ data.

The best fit values of the explosion energy, ejecta mass and ISM density are comparable to those found by \cite{Gelfand2015}, though \cite{Gelfand2015} found that there are degeneracies between some model parameters, and a full exploration of the model parameter space is left for future work.
We also find a small value of wind magnetization parameter $\eta_{\rm B} \sim 10^{-3}$, comparable to other PWNe.
Interestingly, we found that G54.1's particle spectrum is roughly consistent with a single particle index, with the two particle indices $p_1$ and $p_2$ differing by only 0.13.
Most PWNe require two particle spectral indices differing by $\sim 0.5$ to simultaneously account for their X-ray and radio spectra.
This suggests that there may be a diversity of particle acceleration mechanisms operating in PWNe.

We find that an $\approx 2,750$~K photon field is required for our model to reproduce the $\gamma$-ray data points.  
G54.1$+$0.3 is embedded within a cluster of OB stars, with spectra much hotter than $\approx 2,750$~K.
A photon field resembling a blackbody with an $\approx 2,750$~K temperature may be produced after the light from these hot OB stars is reprocessed by the dust known to surround G54.1$+$0.3.

This PWN model predicts a magnetic field $B \approx 7 \mu$G and a PWN age of 2830~yr, which corresponds to a break energy of $\approx 8$~keV (see equations 1 and 2).
This is remarkably close to the value of the break observed at $\approx5$~keV.
If the spectral break observed at 5~keV is due to synchrotron cooling, then this suggests that our SED modeling has correctly predicted the PWN magnetic field strength and age.

\subsection{Comparison with Other Young PWNe}

Table \ref{tab:comparison} lists some of the properties of the G54.1+0.3 system, derived in this paper and in previous studies, with the properties of other young ($\lesssim 5$~kyr) PWNe.
Table \ref{tab:comparison} list these PWNe in order of decreasing spin down power, and demonstates that \src \ is not particularly powerful for its age.
\srcp \ has a relatively long spin down timescale $\tau_{\rm sd}$, comparable to G21.5-0.9, and much longer than Kes 75 and HESS J1640$-$465.
The PWN in G54.1$+$0.3 also has the low PWN magnetization parameter $\eta_{B}$, while \srcp's dipole magnetic field is comparable to the other systems.
\srcp's  spin down luminosity it also typical among these other young PWNe.

Using the spin down timescale and braking index in Table \ref{tab:model_pars}, we have calculated the pulsar's initial spin period $P_0$ by setting the time $t$ equal to the pulsar true age $\tau_{\rm true} \approx 2800$~yr:
\begin{equation}
P_{0} =  P \left( 1 + \frac{t}{\tau_{\rm sd}}     \right)^{\frac{1}{1-n}} = 72~{\rm ms},   
\end{equation}
and initial spin period derivative $\dot{P_0}$:
\begin{equation}
\dot{P_{0}} =  \frac{P_0}{\tau_{\rm sd}(p - 1)} = 7.1 \times 10^{-13} \rm{s \ s^{-1}}.
\end{equation}
These parameters correspond to an initial spin down power initial $\dot{E_0} = 7.5 \times 10^{37}$~erg~s$^{-1}$, and initial spin down measured dipole field $B_0 = 7.2 \times 10^{12} $~G.
We see that the spin down power of Kes 75 and \src \ have decreased from their initial values much less than the decrease inferred for HESS J1640$-$465 and G21.5-0.9.
A future measurement of \srcp's braking index would allow for a more detailed comparison of \src \ with other young PWNe.

Previous \nustar \ studies of young PWNe such as the Crab and 3C 58 found that the PWN size shrinks with increasing energy \citep{an2019,mad2015}.
This 'synchrotron burnoff' effect probes particle transport within these PWNe, and may also be important to understand the structure of \src.
It is unclear if there is a significant 'synchrotron burnoff' in the case of PWN \src, the \nustar \ PSF is  unfortunately comparable to the apparent size of the PWN,  (demonstrating the need for hard X-ray observatories with higher spatial resolution \citep{rey2023}).

\begin{center}
\begin{deluxetable*}{lcccccccccc}
\label{tab:comparison}
\tablewidth{0pt} 
\tablecaption{Young Pulsars Powering PWNe}
\tablehead{
\colhead{Pulsar} & \colhead{SNR} & \colhead{P} & \colhead{$\dot{P}$} & \colhead{$\dot{E}$} & \colhead{$\tau_{\rm char.}$} & \colhead{$B$} & \colhead{$P_0$} & \colhead{$\eta_B$} & \colhead{$E_{\rm max}$} & \colhead{Ref.} \\
\colhead{}            & \colhead{}         & \colhead{(ms)}   & \colhead{($10^{-14}$ s/s)} & \colhead{($10^{36}$ erg/s)} & \colhead{(kyr)}            & \colhead{($10^{12}$ G)}       & \colhead{(ms)} & \colhead{} & \colhead{(PeV)} & \colhead{}
}
\startdata
PSR B0531$+$21     & Crab  & 33  & 42.1   & 450  & 1.2       & 3.8  & $\approx$ 20  &  &  & 1 \\
PSR B0540$-$69     & SNR 0540$-$69.3    & 51  & 47.9   & 150  & 1.7      & 4.9  & $\approx$ 40  &  &  & 1 \\
PSR J1833$-$1034   & G21.5$-$0.9        & 62  & 20.2   & 33   & 4.8        & 3.2  & $\approx$ 20  & $3.2 \times 10^{-3}$  & 0.3  & 2 \\
PSR J0205+6449   & 3C 58            & 66  & 19.3   & 27   & 5.4       & 2.9  & $\approx$ 50  &  &  & 1 \\
PSR B1509$-$58     & MSH 15$-$52        & 152 & 153    & 18   & 1.6        & 15   & $\approx$ 10  &  &  & 1 \\
PSR J1124$-$5916   & G292.0+1.8       & 135 & 76.4   & 12   & 2.9        & 4.3  & $\approx$ 40  &  &  & 1 \\
PSR J1930+1852  & G54.1+0.3        & 137 & 75.1   & 12   & 2.9        & 4.3  &  $\approx$ 72  & 2.4$\times$ $10^{-3}$  & 0.4  & This work \\
PSR J1846$-$0258   & Kes 75           & 327 & 710    & 8.1  & 0.73      & 50   & $\approx$ 200  & 0.12  & 1.1  & 3 \\
PSR J1640$-$4631  & G338.3$-$0.0 & 206 & 97.6  & 4.4 & 3.4 & 14 &  $\approx$ 10 & 0.07 & 1.24 & 4  \\
\enddata
\tablecomments{$P$ and $\dot{P}$ values are taken from the ATNF pulsar catalog.  \\
References:
1  \citep{che2005}
2 \citep{Hattori2020}
3 (\citealt{got2021}; \citealt{Straal2023})
4 (\citealt{Gotthelf2014}; \citealt{abd2023})
}
\end{deluxetable*}
\end{center}

\section{Summary}
\label{summary}
We have analyzed spectral and timing data from previously unpublished \nustar \ observations of PWN \src \ powered by PSR J1930$+$1852.
PWN \src \ is clearly detected up to $\approx 70$~keV, with spectral curvature in the 3$-$30~keV band.
Modeling the PWN SED with the benefit of this  \nustar \ data, and also new radio and $\gamma$~ray data, suggests that the maximum particle energy $E_{\rm max} \sim 400$~TeV.
A future pulsar timing analysis and exploration of the PWN model parameter space can better constrain parameters such as the braking index $n$, the spin down timescale $\tau_{\rm sd}$, and the initial pulsar spin period $P_{0}$.

\begin{acknowledgements}
We acknowledge support from \nustar \ Grant F8732 and NYUAD research grant AD022.
CASS acknowledges support from NYUAD Institute, which is funded by the government of Abu Dhabi through administration of Tamkeen.
GZ acknowledges support from the China Manned Space Program with grant No. CMS-CSST-2025-A13.
\end{acknowledgements}

\end{document}